\documentclass[submission,copyright,creativecommons]{eptcs}
\providecommand{\event}{HaTT 2016} 
\usepackage{breakurl}             

\usepackage{amsmath}
\usepackage{amsfonts}
\usepackage{mathptmx}

\usepackage{proof}
\usepackage{ucs}
\usepackage[utf8x]{inputenc}
\usepackage[T1]{fontenc}

\newcommand{\proves}{\ensuremath{\vdash}}

\newcommand{\Type}{\ensuremath{\mathrm{Type}}}
\newcommand{\Prop}{\ensuremath{\mathrm{Prop}}}

\makeatletter

\newcommand{\Rmnum}[1]{\expandafter\@slowromancap\romannumeral #1@}
\makeatother

\newcommand{\Set}{\mathrm{Set}}

\newcommand{\tof}{\mathcal{F}}
\newcommand{\tog}{\mathcal{G}}
\newcommand{\conv}{\mathcal{C}}

\newcommand{\FV}{\mathrm{FV}}
\newcommand{\FC}{\mathrm{FC}}
\newcommand{\eqvl}{\leftrightarrow}
\newcommand{\case}{\mathtt{case}}

\usepackage{booktabs}

\usepackage{color}
\usepackage[dvipsnames]{xcolor}
\definecolor{darkblue}{HTML}{00007F}

\allowdisplaybreaks

\DeclareSymbolFont{letters}{OML}{txmi}{m}{it}

\title{Goal Translation for a Hammer for Coq (Extended Abstract)}
\author{{\L}ukasz Czajka
\email{lukasz.czajka@uibk.ac.at}
\institute{University of Innsbruck, Austria}
\and
Cezary Kaliszyk
\email{cezary.kaliszyk@uibk.ac.at}
\institute{University of Innsbruck, Austria}}
\def\titlerunning{Goal Translation for a Hammer for Coq (Extended Abstract)}
\def\authorrunning{{\L}. Czajka \& C. Kaliszyk}

\begin{document}
\maketitle

\begin{abstract}
  Hammers are tools that provide general purpose automation for formal
  proof assistants. Despite the gaining popularity of the more
  advanced versions of type theory, there are no hammers for such
  systems. We present an extension of the various hammer components to
  type theory: (i) a translation of a significant part of the Coq
  logic into the format of automated proof systems; (ii) a~proof
  reconstruction mechanism based on a Ben-Yelles-type algorithm
  combined with limited rewriting, congruence closure and a
  first-order generalization of the left rules of Dyckhoff’s
  system~LJT.
\end{abstract}

\section{Introduction}

Justifying small proof steps is usually a significant part of the
process of formalizing proofs in an \emph{interactive theorem proving}
(ITP), or \emph{proof assistant}, system. Many of such goals would be
considered trivial by mathematicians. Still, state-of-the-art~ITPs
require the user to spend an important part of the formalization
effort on them. The main points that constitute this effort are
usually library search, minor transformations on the already proved
theorems (such as reordering assumptions or reasoning modulo
associativity-commutativity), as well as combining a small number of
simple known lemmas. To reduce this effort various automation
techniques have been conceived, including techniques from automated
reasoning and domain specific decision procedures. The strongest
general propose automation technique, available for various
interactive theorem provers today is provided by
``hammers''~\cite{jbcklpju-h4qed-jfr}.

Hammers are proof assistant tools that employ external automated
theorem provers (ATPs) in order to automatically find proofs of user
given conjectures. There are three main components of a hammer:
\begin{itemize}
\item Lemma selection (also called relevance filtering or premise
  selection) that heuristically chooses a subset of the accessible
  lemmas that are likely useful for the given conjecture.
\item Translation (encoding) of the user given conjecture together
  with the selected lemmas to the logics and input formats of automated
  theorem provers (ATPs). The focus is usually on first-order logic as
  the majority of the most efficient ATPs today support this
  foundation. The automated systems are in turn used to either find an
  ATP proof or just further narrow down the subset of lemmas to
  precisely those that are necessary in the proof.
\item Proof reconstruction, which uses the obtained information from
  the successful ATP run, to reprove the lemma in the logic of the
  proof assistant.
\end{itemize}

Robust hammers exist for proof assistants based on higher-order logic
(Sledgehammer~\cite{sledgehammer10} for
Isabelle/HOL~\cite{WenzelPN08}, HOLyHammer~\cite{holyhammer} for HOL
Light~\cite{harrison-2009} and HOL4~\cite{slind-norrish-2008}) or
dependently typed set theory (MizAR~\cite{ckjumiz40} for
Mizar~\cite{Bancerek15,Wiedijk2007,Bancerek03}). The general-purpose
automation provided by the most advanced hammers is able to solve
40--50\% of the top-level goals in various
developments~\cite{jbcklpju-h4qed-jfr}, as well as more than 70\% of
the user-visible subgoals~\cite{mash2}, and as such has been found
very useful in various proof developments~\cite{Hales-Developments}.

Despite the gaining popularity of the more advanced versions of type
theory, implemented by systems such as Agda~\cite{BoveDN09Agda},
Coq~\cite{Bertot08Coq}, Lean~\cite{MouraKADR15Lean}, and
Matita~\cite{Matita14}, there are no hammers for such systems.  The
construction of such a tool has so far been hindered by the lack of a
usable encoding component, as well as by comparatively weak proof
reconstuction.

For the proof assistants whose logics are based on the Calculus of
Constructions and its extensions, the existing encodings in
first-order logic so far cover only limited fragments of the source
logic~\cite{GandalfFOL,optimizedencod,BezemHendriksNivelle2002}. Why3~\cite{FilliatreP13}
provides a translation from its own logic~\cite{onelogictorule} (which
is a subset of the Coq logic, including features like rank-1
polymorphism, algebraic data types, recursive functions and inductive
predicates) to the format of various first-order provers (in fact Why3
has been initially used as a translation back-end for
HOLyHammer). Recently, an encoding of the dependently typed
higher-order logic of~$\mathrm{F}^*$ into first-order logic has also
been developed~\cite{f_star}.


The built-in HOL automation is able to reconstruct the majority of the
automatically found proofs using either internal proof
search~\cite{hurd2003d} or source-level reconstruction. The internal
proof search mechanisms provided in Coq, such as the
\texttt{firstorder} tactic~\cite{Corbineau03}, have been insufficient
for this purpose so far. Matita's ordered
paramodulation~\cite{DBLP:conf/mkm/AspertiT07} is able to reconstruct
many goals with up to two or three premises, and the
congruence-closure based internal automation techniques in
Lean~\cite{MouraSelsam2016} are also promising.

The SMTCoq~\cite{smtcoq} project has developed an approach to use
external SAT and SMT solvers and verify their proof witnesses. Small
checkers are implemented using reflection for parts of the SAT and SMT
proof reconstruction, such as one for CNF computation and one for
congruence closure. The procedure is able to handle Coq goals in the
subset of the logic that corresponds to the logics of the input
systems.

\medskip
\noindent{\bf Contributions.} We present our recently developed proof
advice components for type theory and systems based on it. We first
introduce an encoding of the Calculus of Inductive Constructions,
including the additional logical constructions introduced by the Coq
system, in untyped first-order logic with equality. We implement the
translation and evaluate it experimentally on the standard library of
the Coq proof assistant. We advocate that the encoding is sufficient
for a hammer system for Coq: the success rates are comparable to those
demonstrated by early hammer systems for Isabelle/HOL and Mizar, while
the dependencies used in the ATP proofs are most often sufficient to
prove the original theorems. Strictly speaking, our translation is
neither sound nor complete. However, our experiments suggest that the
encoding is ``sound enough'' to be usable. Moreover, we believe that a
``core'' version of the translation is sound and we are currently
working on a proof of this fact.

Secondly, we present a proof reconstruction mechanism based on a
Ben-Yelles-type procedure combined with a first-order generalization
of the left rules of Dyckhoff’s LJT, congruence closure and heuristic
rewriting. With this still preliminary proof search procedure we are
able to reprove almost 90\% of the problems solved by the~ATPs, using
the dependencies extracted from the~ATP output.

\section{Translation}

In this section we introduce an encoding of (a close approximation of)
the Calculus of Inductive Constructions into untyped first-order logic
with equality. The encoding should be a practical one, which implies
that its general theoretical soundness is not the main focus, i.e., of
course the translation needs to be ``sound enough'' to be usable, but
it is more important that the encoding is efficient enough to provide
practically useful information about the necessary proof
dependencies. In particular, the encoding needs to be shallow, meaning
that Coq terms of type~$\Prop$ are translated directly to
corresponding first-order formulas. Our translation is in fact
unsound, e.g., it assumes proof irrelevance and ignores certain
universe constraints. However, we believe that under the assumption of
proof irrelevance a ``core'' version of the translation is sound, and
we are currently working on a proof.

Below we present a variant of the translation for a fragment of the
logic of~Coq. The intention here is to provide a general idea, but not
to describe the encoding in detail. In the first-order language we
assume a unary predicate~$P$, a binary predicate~$T$ and a binary
function symbol~$@$. Usually, we write $t s$ instead of $@(t, s)$.

For the sake of efficiency, terms of type~$\Prop$ are encoded directly
as~FOL formulas using a function~$\tof$. Terms that have type~$\Type$
but not~$\Prop$ are encoded using a function~$\tog$ as guards which
essentially specify what it means for an object to have the given
type. For instance, $\forall f : \tau . \varphi$ where $\tau = \Pi x :
\alpha . \beta$ is translated to $\forall f . \tog(\tau, f) \to
\tof(\varphi)$ where $\tog(\tau, f) = \forall x . \tog(\alpha, x) \to
\tog(\beta, f x)$. So~$\tog(\tau, f)$ says that an object~$f$ has
type~$\tau = \Pi x : \alpha . \beta$ if for any object~$x$ of
type~$\alpha$, the application~$f x$ has type~$\beta$.

\noindent Function~$\tof$ encoding propositions as FOL formulas is
defined by:
\begin{itemize}
\item If $\Gamma \proves t : \Prop$ then $\tof_\Gamma(\Pi x : t . s) =
  \tof_{\Gamma}(t) \to \tof_{\Gamma,x:t}(s)$.
\item If $\Gamma \not\proves t : \Prop$ then $\tof_\Gamma(\Pi x : t
  . s) = \forall x . \tog_{\Gamma}(t, x) \to \tof_{\Gamma,x:t}(s)$.
\item Otherwise, if none of the above apply, $\tof_\Gamma(t) =
  P(\conv_\Gamma(t))$.
\end{itemize}

\noindent Function~$\tog$ encoding types as guards is defined by:
\begin{itemize}
\item If $t = \Pi x : t_1 . t_2$ and $\Gamma \proves t_1 : \Prop$ then
  $\tog_\Gamma(\Pi x : t_1 . t_2, s) = \tof_{\Gamma}(t_1) \to
  \tog_{\Gamma,x:t_1}(t_2, s)$.
\item If $t = \Pi x : t_1 . t_2$ and $\Gamma \not\proves t_1 : \Prop$
  then $\tog_\Gamma(\Pi x : t_1 . t_2, s) = \forall x
  . \tog_{\Gamma}(t_1, x) \to \tog_{\Gamma,x:t_1}(t_2, s x)$.
\item Otherwise, when~$t$ is not a product $\tog_\Gamma(t, s) = T(u,
  \conv_\Gamma(t))$.
\end{itemize}

\noindent Function~$\conv$ encoding terms as FOL terms is defined by:
\begin{itemize}
\item $\conv_\Gamma(b) = b$ for~$b$ being a variable or a constant,
\item $\conv_\Gamma(t s)$ is equal to:
  \begin{itemize}
  \item $\conv_\Gamma(t)$ if $\Gamma \proves s : A : \Prop$ for
    some~$A$,
  \item $\conv_\Gamma(t) \conv_\Gamma(s)$ otherwise.
  \end{itemize}
\item $\conv_\Gamma(\Pi x : t .  s) = P \vec{y}$ for a fresh
  constant~$P$ where $\vec{y} = \FV(\Pi x : t . s)$ and
  \begin{itemize}
  \item if $\Gamma \proves (\Pi x : t . s) : \Prop$ then $\forall
    \vec{y} . P \vec{y} \eqvl \tof_\Gamma(\Pi x : t . s)$ is a new
    axiom,
  \item if $\Gamma \not\proves (\Pi x : t . s) : \Prop$ then $\forall
    \vec{y} z . P \vec{y} z \eqvl \tog_\Gamma(\Pi x : t . s, z)$ is a
    new axiom.
  \end{itemize}
\item $\conv_\Gamma(\lambda \vec{x} : \vec{t} . s) = F \vec{y}$
  where~$s$ does not start with a lambda-abstraction any more, $F$ is
  a fresh constant, $\vec{y} = \FV(\lambda \vec{x} : \vec{t} . s)$ and
  $\forall \vec{y} . \tof_\Gamma(\forall \vec{x} : \vec{t} . F \vec{y}
  \vec{x} = s)$ is a new axiom.
\item $\conv_\Gamma(\case(t, c, n, \lambda \vec{a} : \vec{\alpha}
  . \lambda x : c \vec{p} \vec{a}. \tau, \lambda \vec{x_1} :
  \vec{\tau_1} . s_1, \ldots, \lambda \vec{x_k} : \vec{\tau_k} . s_k))
  = F \vec{y_1} \vec{y_2}$ for a fresh constant~$F$ where
  \begin{itemize}
  \item $I(c : \gamma : \kappa := c_1 : \gamma_1 : \kappa_1, \ldots,
    c_k : \gamma_k : \kappa_k) \in E$,
  \item $\Gamma_2 = \vec{y_2} : \vec{\rho_2} = \FC(\Gamma;t)$,
  \item $\Gamma_1 = \vec{y_1} : \vec{\rho_1} = \FC(\Gamma;\lambda
    \vec{y_2} : \vec{\rho_2} . t (\lambda \vec{x_1} : \vec{\tau_1}
    . s_1) \ldots (\lambda \vec{x_k} : \vec{\tau_k} . s_k))$,
  \item $\gamma_i = \Pi \vec{z_i} : \vec{\beta_i} . \Pi \vec{x_i} :
    \vec{\tau_i} . \sigma_i$ for $i=1,\ldots,k$,
  \item the following is a new axiom:
    \[
    \begin{array}{rcl}
      \forall \vec{y_1} . \tof_{\Gamma_1}(\forall \vec{y_2} : \vec{\rho_2}
      &.& (\exists \vec{z_1} : \vec{\beta_1} . \exists \vec{x_1} : \vec{\tau_1} . t = c_1 \vec{z_1} \vec{x_1} \land F
      \vec{y_1} \vec{y_2} = s_1) \\
      &\lor& \ldots \\
      &\lor& (\exists \vec{z_k} : \vec{\beta_k} . \exists \vec{x_k} : \vec{\tau_k} . t = c_k \vec{z_k} \vec{x_k} \land F
      \vec{y_1} \vec{y_2} = s_k))
    \end{array}
    \]
  \end{itemize}

  Here $t$ is the term matched on, the type of~$t$ has the form $c
  \vec{p} \vec{u}$, the integer~$n$ denotes the number of parameters
  (which is the length of~$\vec{p}$), the
  type~$\tau[\vec{u}/\vec{a},t/x]$ is the return type, i.e., the type
  of the whole case expression, $\vec{a} \cap \FV(\vec{p}) =
  \emptyset$, and~$s_i[\vec{v}/\vec{x_i}]$ is the value of the case
  expression if the value of~$t$ is~$c_i \vec{p} \vec{v}$. The free
  variable context~$\FC(\Gamma;t)$ of~$t$ in~$\Gamma$ is defined
  inductively: $\FC(\emptyset; t) = \emptyset$; $\FC(\Gamma, x : \tau;
  t) = \FC(\Gamma; \lambda x : \tau . t), x : \tau$ if $x \in \FV(t)$;
  and $\FC(\Gamma, x : \tau; t) = \FC(\Gamma; t)$ if $x \notin
  \FV(t)$.
\end{itemize}

In the data exported from Coq there are three types of declarations:
definitions, typing declarations and inductive declarations. We
briefly describe how all of them are translated.

A definition $c = t : \tau : \kappa$ is translated as follows.
\begin{itemize}
\item If $\kappa = \Prop$ then add $\tof(\tau)$ as a new axiom with
  label~$c$.
\item If $\kappa \ne \Prop$ then
  \begin{itemize}
  \item add $\tog(\tau, c)$ as a new axiom,
  \item if $\tau = \Prop$ then add $c \eqvl \tof(t)$ as a new axiom
    with label~$c$,
  \item if $\tau = \Set$ or $\tau = \Type$ then add $\forall f . c f
    \eqvl \tog(t, f)$ as a new axiom with label~$c$,
  \item if $\tau \notin \{\Prop,\Set,\Type\}$ then add the equation $c
    = \conv(t)$ as a new axiom with label~$c$.
  \end{itemize}
\end{itemize}

A typing declaration $c : \tau : \kappa$ is translated as follows.
\begin{itemize}
\item If $\kappa = \Prop$ then add $\tof(\tau)$ as a new axiom with
  label~$c$.
\item If $\kappa \ne \Prop$ then add $\tog(\tau, c)$ as a new axiom
  with label~$c$.
\end{itemize}

An inductive declaration $I(c : \tau : \kappa := c_1 : \tau_1 :
\kappa_1, \ldots, c_n : \tau_n : \kappa_n)$ is translated as follows.
\begin{itemize}
\item Translate the typing declaration $c : \tau : \kappa$.
\item Translate each typing declaration $c_i : \tau_i : \kappa$ for
  $i=1,\ldots,n$.
\item Add axioms stating injectivity of constructors, axioms stating
  non-equality of different constructors, and the ``inversion'' axioms
  for elements of the inductive type.
\end{itemize}

For inductive types also induction principles and recursor definitions
are translated.

\smallskip

The above only gives a general outline of the translation. In
practice, we make a number of optimisations, e.g., the arity
optimisation by Meng and Paulson~\cite{MengPaulson2008}, or
translating fully applied functions with target type~Prop directly to
first-order predicates.

\section{Reconstruction}

We report on our work on proof reconstruction. We evaluate the Coq
internal reconstruction mechanisms including \texttt{tauto} and
\texttt{firstorder}~\cite{Corbineau03} on the original proof
dependencies and on the ATP found proofs, which are in certain cases
more precise. In particular \texttt{firstorder} seems insufficient for
finding proofs for problems created using the advice obtained from
the~ATP runs. This is partly caused by the fact that it does not fully
axiomatize equality, but even on problems which require only purely
logical first-order reasoning its running time is sometimes
unacceptable.

The formulas that we attempt to reprove usually belong to fragments of
intuitionistic logic low in the Mints hierarchy~\cite{urzy1}. Most of
proved theorems follow by combining a few known lemmas. This raises a
possibility of devising an automated proof procedure optimized for
these fragments of intuitionistic logic, and for the usage of the
advice obtained from the~ATP runs. We implemented a preliminary
version of a Ben-Yelles-type procedure (essentially
\texttt{eauto}-type proof search with a looping check) augmented with
a first-order generalization of the left rules of Dyckhoff's system
LJT~\cite{Dyckhoff1992}, the use of the~\texttt{congruence} tactic,
and heuristic rewriting using equational hypotheses.

It is important to note that while the external~ATPs we employ are
classical and the translation assumes proof irrelevance, the proof
reconstruction phase does not assume any additional axioms. We reprove
the theorems in the intuitionistic logic of Coq, effectively using the
output of the~ATPs merely as hints for our hand-crafted proof search
procedure. Therefore, if the~ATP proof is inherently classical then
proof reconstruction will fail. Currently, the only information
from~ATP runs we use is a list of lemmas needed by the~ATP to prove
the theorem (these are added to the context) and a list of constant
definitions used in the~ATP proof (we try unfolding these constants
and no others).

Another thing to note is that we do not use the information contained
in the Coq standard library during reconstruction. This would not make
sense for our evaluation of the reconstruction mechanism, since we try
to reprove the theorems from the Coq standard library. In particular,
we do not use any preexisting hint databases available in Coq, not
even the core database (we use the \texttt{auto} and \texttt{eauto}
tactics with the \texttt{nocore} option). Also, we do not use any
domain-specific decision procedures available as~Coq tactics, e.g.,
\texttt{field}, \texttt{ring} or \texttt{omega}.

\section{Evaluation}

We evaluated our translation on the problems generated from all
declarations of terms of type~$\Prop$ in the Coq standard library of
Coq version~8.5. We used the following classical~ATPs: E Prover
version~1.9~\cite{Schulz13}, Vampire version~4.0~\cite{Vampire} and Z3
version~4.0~\cite{de-moura-bjoerner-2008}. The methodology was to
measure the number of theorems that the~ATP could reprove from their
extended dependencies within a time limit of~30~s for each
problem. The extended dependencies of a theorem are obtained by taking
all constants occuring in the proof term of the theorem in Coq
standard library, and recursively taking all constants occuring in the
types and non-proof definitions of any dependencies extracted so
far. Because of the use of extended dependencies, the average number
of generated~FOL axioms for a problem is~193. We limited the recursive
extraction of extended dependencies to depth~2.

The evaluation was performed on a 48-core server with 2.2~GHz~AMD
Opteron CPUs and 320~GB RAM. Each problem was always assigned one CPU
core. Table~\ref{tab_1} shows the results of our evaluation. The
column ``Solved\%'' denotes the percentage (rounded to the first
decimal place) of the problems solved by a given prover, and
``Solved'' the number of problems solved out of the total number
of~20803 problems. The column ``Sum\%'' denotes the percentage, and
``Sum'' the total number, of problems solved by the prover or any of
the provers listed above it. The column ``Unique'' denotes the number
of problems the given prover solved but no other prover could solve.

\begin{table*}[t]
\centering \setlength{\tabcolsep}{3mm}
  \begin{tabular}[c]{lccccc}\toprule
    Prover & Solved\% & Solved & Sum\% & Sum & Unique
    \\
    \midrule
    Vampire & 32.9 & 6839 & 32.9 & 6839 & 855 \\
    Z3 & 27.6 & 5734 & 34.9 & 7265 & 390\\
    E Prover & 25.8 & 5376 & 35.3 & 7337 & 72\\\midrule
    any & 35.3 & 7337 & 35.3 & 7337 & \\
    \bottomrule
  \end{tabular}\vspace{1mm}
  \caption{Results of the experimental evaluation on the 20803 FOL
    problems generated from the propositions in the Coq standard
    library.}\label{tab_1}
\end{table*}

We also evaluated various proof reconstruction mechanisms on the
problems originating from ATP proofs of lemmas in the Coq standard
library. In our setting, the Ben-Yelles-type algorithm mentioned in
the previous section tends to perform significantly better than the
available Coq's tactics. The results of the evaluation are presented
in Table~\ref{tab_2}. Our tactic (\texttt{yreconstr}) manages to
reconstruct about 88\% of the reproved theorems. However, it needs to
be remarked that if we use the advice obtained from ATP runs then
about 50\% of the the reproved theorems follow by a combination of
hypothesis simplification, the tactics \texttt{intuition},
\texttt{auto}, \texttt{easy}, \texttt{congruence} and a few heuristics
(tactic \texttt{simple}). Moreover, the \texttt{yreconstr} tactic
without any hints (\texttt{yreconstr0}), i.e., without using any of
the information obtained from ATP runs, achieves a success rate of
about 26\%. The reconstruction success rate of the \texttt{firstorder}
tactic combined with various heuristics is about 70\% if generic
axioms for equality are added to the context (tactic
\texttt{firstorder'}). The \texttt{jp} tactic (which integrates the
intuitionistic first-order automated theorem prover
JProver~\cite{DBLP:conf/cade/SchmittLKN01} into Coq) combined with
various heuristics and equality axioms (tactic \texttt{jprover})
achieves a reconstruction success rate of about 56\%. This low success
rate is explained by the fact that in contrast to the
\texttt{firstorder} tactic the~\texttt{jp} tactic cannot be
parameterised by a tactic used at the leaves of the search tree when
no logical rule applies.

\begin{table*}[t]
  \centering \setlength{\tabcolsep}{3mm}
  \begin{tabular}[c]{lccc}\toprule
    Tactic & Time & Solved\% & Solved \\
    \midrule
    \texttt{yreconstr0} & 10s & 26.8 & 1965 \\
    \texttt{yreconstr} & 1s & 83.1 & 6097 \\
    \texttt{yreconstr} & 2s & 85.8 & 6296 \\
    \texttt{yreconstr} & 5s & 87.5 & 6421 \\
    \texttt{yreconstr} & 10s & 88.1 & 6466 \\
    \texttt{yreconstr} & 15s & 88.2 & 6473 \\
    \texttt{simple} & 1s & 50.1 & 3674 \\
    \texttt{firstorder'} & 10s & 69.6 & 5103 \\
    \texttt{jprover} & 10s & 56.1 & 4114 \\\midrule
    any &  & 90.1 & 6609 \\
    \bottomrule
  \end{tabular}\vspace{1mm}
  \caption{Results of the evaluation of proof reconstruction on the
    7337 problems solved by the~ATPs.}\label{tab_2}
\end{table*}

\medskip
\noindent {\bf Acknowledgments.} We thank the organizers of the First
Coq Coding Sprint, especially Yves Bertot, for the help with
implementing Coq export plugins. We wish to thank Thibault Gauthier
for the first version of the Coq exported data, as as well as Claudio
Sacerdoti-Coen for improvements to the exported data and fruitful
discussions on Coq proof reconstruction. This work has been supported
by the Austrian Science Fund (FWF) grant P26201.

\bibliographystyle{eptcs}
\bibliography{generic}

\end{document}